\documentclass[journal=jctcce,manuscript=article,layout=twocolumn]{achemso}

\usepackage{algorithm}
\usepackage{algpseudocode}
\usepackage[version=4]{mhchem}
\usepackage{amsfonts}
\usepackage{amssymb,amsmath}
\usepackage{amsthm}
\usepackage{enumerate}
\usepackage{bbm}
\usepackage{graphicx}
\usepackage{bm}
\usepackage{siunitx}
\DeclareSIUnit\angstrom{\text{Å}}
\usepackage{mathtools}
\usepackage{diffcoeff}
\usepackage{ulem}
\usepackage{subcaption}
\usepackage{booktabs}
\usepackage{multirow}
\usepackage{placeins}
\usepackage{upgreek}
\usepackage[export]{adjustbox}

\newtheorem{theorem}{Theorem}

\usepackage{hyperref}

\usepackage{cleveref}
\crefname{equation}{Eq.}{Eqs.}
\Crefname{equation}{Eq.}{Eqs.}
\crefname{figure}{Fig.}{Figs.}
\Crefname{figure}{Fig.}{Figs.}
\crefname{table}{Table}{Tables}
\Crefname{table}{Table}{Tables}
\crefname{section}{Section}{Sections}
\Crefname{section}{Section}{Sections}
\crefname{subsection}{Section}{Sections}
\Crefname{subsection}{Section}{Sections}
\crefname{subsubsection}{Section}{Sections}
\Crefname{subsubsection}{Section}{Sections}
\crefname{theorem}{Theorem}{Theorems}
\Crefname{theorem}{Theorem}{Theorems}
\crefname{claim}{Claim}{Claims}
\Crefname{claim}{Claim}{Claims}
\crefname{algorithm}{Algorithm}{Algorithms}
\Crefname{algorithm}{Algorithm}{Algorithms}
\creflabelformat{equation}{(#2#1#3)}

\usepackage{etoolbox}
\pretocmd{\cite}{~}{}{}

\makeatletter
\newcommand{\WidenSITocSectionNumbers}{%
  \patchcmd{\l@section}{\setlength\@tempdima{1.5em}}%
    {\setlength\@tempdima{2.5em}}{}%
    {\PackageWarning{SI}{Could not widen section numbers in the contents}}%
}
\makeatother

\let\Psi\varPsi
\let\Phi\varPhi
\let\Theta\varTheta
\let\Omega\varOmega

\newcommand{\jux}{\textendash}

\newcommand{\Schrodinger}{Schr\"{o}dinger }
\newcommand{\vac}{\textrm{vac}}

\newcommand{\ket}[1]{|#1 \rangle}

\newcommand{\mel}[3]{\langle #1|#2 \ket{#3}}

\newsavebox{\OtwoPanelBox}
\newcommand{\OtwoPanel}[2]{%
	\sbox{\OtwoPanelBox}{\includegraphics[height=0.488\textwidth]{#2}}%
	\makebox[0pt][l]{\hspace*{\dimexpr\wd\OtwoPanelBox/2\relax}%
		\raisebox{\dimexpr\ht\OtwoPanelBox/2\relax}[0pt][0pt]{\hypertarget{#1}{}}}%
	\usebox{\OtwoPanelBox}%
}

\newcommand{\TableRuleSpacing}{%
  \let\OriginalHline\hline
  \def\hhline{%
    \noalign{\vskip 2pt}%
    \OriginalHline
    \noalign{\vskip\doublerulesep}%
    \OriginalHline
    \noalign{\vskip 2pt}%
  }%
  \def\hline{%
    \noalign{\vskip 2pt}%
    \OriginalHline
    \noalign{\vskip 2pt}%
  }%
}

\usepackage{xcolor}
\usepackage{letltxmacro,xparse}

\DeclareRobustCommand{\lyz}[2]{%
  \texorpdfstring{%
    \begingroup
    \LetLtxMacro{\lyzOriginalcref}{\cref}%
    \LetLtxMacro{\lyzOriginalCref}{\Cref}%
    \RenewDocumentCommand{\cref}{s m}{%
      \mbox{\IfBooleanTF{##1}{\lyzOriginalcref*{##2}}{\lyzOriginalcref{##2}}}%
    }%
    \RenewDocumentCommand{\Cref}{s m}{%
      \mbox{\IfBooleanTF{##1}{\lyzOriginalCref*{##2}}{\lyzOriginalCref{##2}}}%
    }%
    \textcolor{blue}{\sout{#1}}%
    \endgroup
    \space\textcolor{blue}{#2}%
  }{#2}%
}

\newcommand{\BNUChemAffiliation}{Key Laboratory of Theoretical and Computational Photochemistry, Ministry of Education, College of Chemistry, Beijing Normal University, Beijing, 100875, China}
\newcommand{\BNUStudyAffiliation}{Institute for Advanced Study, Beijing Normal University, Beijing, 100875, China}
\newcommand{\BNUAuthor}[1]{%
  \author{#1}%
  \affiliation{\BNUChemAffiliation}%
  \alsoaffiliation{\BNUStudyAffiliation}%
}
\newcommand{\BNUCorrespondingAuthor}[2]{%
  \BNUAuthor{#1}%
  \email{#2}%
}
\newcommand{\BNUManuscriptAuthors}{%
  \BNUAuthor{Bohan Zhang}%
  \BNUAuthor{Zibo Wu}%
  \BNUCorrespondingAuthor{Yunzhi Li}{dmzliyunzhi@bnu.edu.cn}%
  \BNUAuthor{Wei-Hai Fang}%
  \BNUCorrespondingAuthor{Zhendong Li}{zhendongli@bnu.edu.cn}%
}
\newcommand{\ManuscriptTitle}{%
Spin-Adapted Neural-Network Backflow Wavefunctions for Excited States
of Strongly Correlated Open-Shell Systems%
}
\newcommand{\ManuscriptShortTitle}{SA-NNBF+NES}

\title[\ManuscriptShortTitle]
{\ManuscriptTitle}

\BNUManuscriptAuthors

\begin{document}

\begin{abstract}
Building on the success of neural-network quantum states (NQS) for molecular ground states, multi-state frameworks such as natural excited states (NES) [Pfau et al., \textit{Science} \textbf{2024}, \textit{385}, eadn0137] extend NQS to chemically important excited states.
However, without control over total spin, such calculations can require the inclusion of many intervening states of other spins, particularly in strongly correlated systems with dense low-energy spectra.
We therefore generalize our previously developed spin-adapted neural network backflow (SA-NNBF) ansatz [Li et al., arXiv:2604.06841 (2026)] to improve its variational flexibility for excited-state calculations within a prescribed spin sector.
We establish the universality of this extended ansatz within the chosen orbital space and spin sector, and use numerical benchmarks to assess how parameter allocation affects the ansatz's accuracy.
With the spin-adapted ansatz, 
we are able to directly optimize the first excited state (a singlet) of \ce{O2} without using any multi-state frameworks.
Combining the extended ansatz with NES, we obtain low-lying triplet states of \ce{O2} and singlet states of the \ce{[Fe2S2(SCH3)4]^{2-}} cluster without including intervening states of other spins.
Both the single-state and NES calculations achieve chemical accuracy relative to the exact energies.
Our work provides a practical NQS framework for targeting excited states of prescribed spin in strongly correlated systems.
\end{abstract}

\section{Introduction}\label{sec:Introduction}

Accurate calculations of molecular excited states are important for understanding electronic spectra and photochemical processes, but remain especially challenging in strongly correlated systems.
Time-dependent density functional theory (TD-DFT)\cite{1995-Casida-TimeDependent} is most commonly used in practical applications for its low computational cost with reasonable accuracy for many excitations.
However, the performance of a TD-DFT calculation depends on the choice of exchange\jux correlation functional and the character of the excitation \cite{2005-Dreuw-SingleReference, 2022-Liang-Revisiting}, which sometimes makes its results unreliable, especially for strongly correlated systems.
Wavefunction methods, on the other hand, give overall more precise results with different ways to represent the many-body wavefunction, such as equation-of-motion coupled-cluster theory with single and double excitations (EOM-CCSD)\cite{2009-Shavitt-ManyBody, 2021-Gonzalez-Quantum}, 
 
 selected configuration interaction (SCI)\cite{2020-Tubman-Modern, 2016-Holmes-HeatBath, 2016-Liu-ICIa, 2020-Zhang-Iterative, 2021-Zhang-Further} 
 
and density matrix renormalization group (DMRG)\cite{2015-Szalay-Tensor, 2020-Baiardi-Density}
while these methods are usually more expensive and each of them has its specific limitations\cite{2008-Krylov-Equationofmotion, 2021-Zhang-Further, 2016-Holmes-HeatBath, 2020-Tubman-Modern, 2020-Baiardi-Density, 2016-Chan-Matrix}.

Recently, neural-network quantum states (NQS) have provided a new way to represent the wavefunction using the expressive power of neural networks\cite{2017-Carleo-Solving, 2020-Hibat-Allah-Recurrent, 2023-Wu-Tensornetwork}.
In recent NQS studies, one of the most popular ans\"{a}tze for fermions is the neural network backflow (NNBF), 
which uses determinants produced by a neural network to describe fermionic wavefunctions \cite{2019-Luo-Backflowb, 2024-Liu-Neuralb, 2024-Liu-Unifying}.
Related ans\"{a}tze introduce additional degrees of freedom, as in the hidden fermion determinant state (HFDS)\cite{2022-RobledoMoreno-Fermionic}.
Others replace the determinant with a Pfaffian, as in the hidden fermion Pfaffian state (HFPS)\cite{2025-Chen-Neural, 2025-Roth-Superconductivity}.
To solve for the ground state of a system, an NQS is usually optimized through variational Monte Carlo (VMC) where the energy and gradients are estimated by stochastic sampling and optimized by iteratively updating the trainable parameters in the wavefunction ansatz\cite{2017-Becca},
with a favorable low scaling with the system size while retaining a high accuracy for correlated states.

To extend NQS-VMC to excited states, a straightforward way is to
optimize separate NQS sequentially or jointly with overlap penalties, but the penalty strengths require tuning and sampled overlaps can be difficult to estimate\cite{2021-Pathak-Excited, 2023-Entwistle-Electronic, 2023-Liu-Calculate, 2024-Wheeler-OverlapExcited, 2024-Szabo-Improved, 2026-Gao-ExcitedPfaffians}.
The auxiliary wavefunction approach avoids penalty tuning by projecting out lower states, but still requires their prior optimization and overlap estimates\cite{2023-Lu-Penalty}.
To avoid explicit overlap estimates and penalty parameters,
the natural excited states (NES) method recasts several low-lying states as a ground-state variational problem in an extended configuration space\cite{2024-Pfau-NES, 2025-Ma-NES, 2025-Hendry-GrassmannNES},
where the energies of individual states then follow from diagonalizing the sampled average of the matrix local energy.

\begin{figure}[t]
	\centering
	\includegraphics[width=0.9\linewidth]{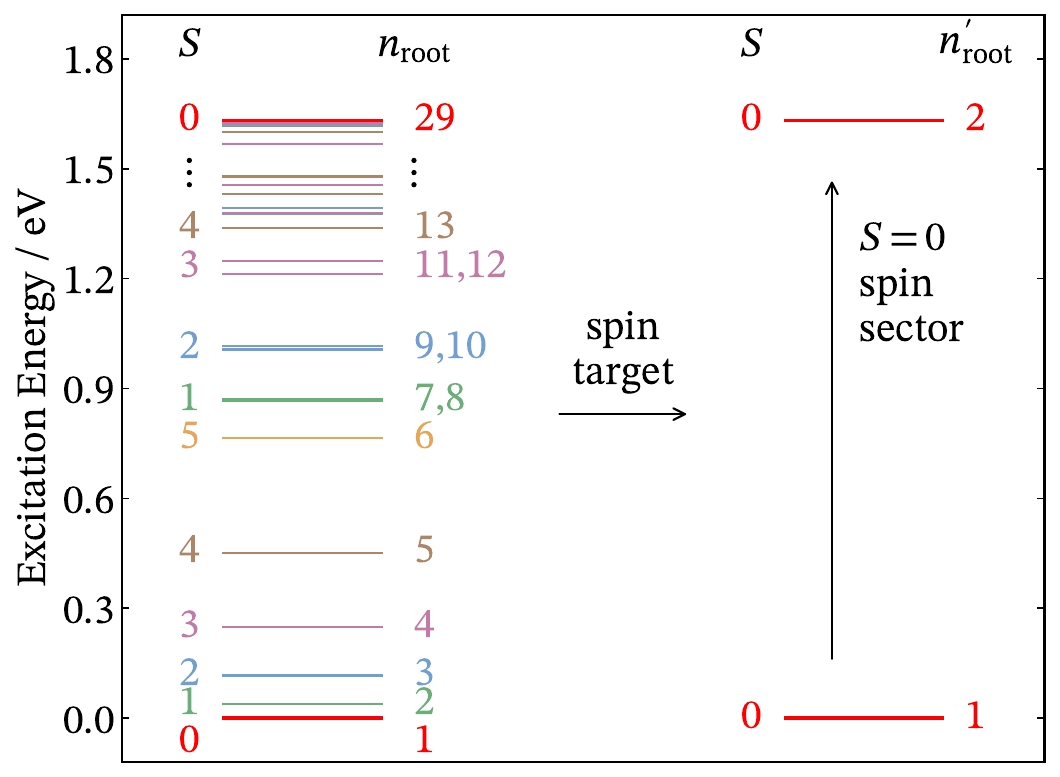}
	\caption{FCI excitation energies of [2Fe-2S] relative to its $S=0$ ground state. Each horizontal line represents a state in the $M_{\rm s}=0$ space; colors indicate total spin, with $S=0$ in red. The left spectrum includes all spin sectors through the first excited singlet, while the right spectrum retains only the targeted $(S,M_{\rm s})=(0,0)$ sector. The labels $n_{\rm root}$ and $n_{\rm root}'$ give energy-order indices in the full and $S=0$ sector, respectively.}
	\label{fig:fe2s2-fci}
\end{figure}

Although the methods above provide access to low-lying excited states, 
solving for a specific target state with such methods requires including and optimizing all the states below the target one.
This could make the calculation inefficient, especially for strongly correlated systems where many near-degenerate states with different spins intervene.
For example, for the active space model of \ce{[Fe2S2(SCH3)4]^{2-}} cluster (hereafter [2Fe-2S])\cite{2017-Li-SPMPS}, the first excited singlet is the \num{29}-th state in the $M_{\rm s}=0$ spectrum. The \num{28} lower-energy states, including the singlet ground state, must also be optimized to reach it with the methods above; see \cref{fig:fe2s2-fci} for an illustration. 
To reduce the number of states needed to be included, a well-established strategy which is already exploited in electronic structure methods, is to restrict the optimization in the target-spin sector to exclude the intervening states of other spins\cite{2012-Sharma-SpinAdapted, 2014-Wouters-CheMPS2, 2014-Sharma-Low, 2019-Dobrautz-Efficient, 2017-Fales-Robust, 2022-Weser-Spin, 2022-Mejuto-Zaera-Effect}.
For NQS-VMC, a notable attempt is using spin penalties that bias excited-state NQS calculations toward a target total spin\cite{2024-Szabo-Improved}, but the coefficients of the penalties require tuning and the 
purity of the target spin depends on the optimization.

Recently, inspired by the spin-adapted antisymmetrization method (SAAM) in real space \cite{2025-Li-SAAM}, we developed a spin-adapted NQS ansatz in occupied-number representation, namely the spin-adapted neural network backflow (SA-NNBF), which yields strictly spin-pure wavefunctions by construction\cite{2026-Li-SANNBF}.
In this work, we further extend SA-NNBF to a multi-determinant form and prove that an SA-NNBF is flexible enough to represent any state in the chosen orbital space and spin sector with an arbitrary fixed nonzero spin function.
Moreover, we combine it with NES to optimize multiple low-lying states directly within fixed spin sectors, without any additional spin penalties.
With spin-adaptation, the example of the 29-state calculation in [2Fe-2S] could be reduced to only 2 states as shown in \cref{fig:fe2s2-fci}.
We test such strategy on the example of \ce{O2} and [2Fe-2S],
where several low-lying states of specific spins are obtained without additional states included in the NES calculation, 
and the energies of the resulting states are within chemical accuracy.
These results demonstrate that the combination of SA-NNBF and NES provides an efficient tool to calculate excited states with high accuracy even for strongly correlated systems.

The remaining parts of this work are organized as follows.
\Cref{sec:Theory} reviews variational Monte Carlo with neural quantum states, the natural excited states method, and the NNBF and SA-NNBF ans\"{a}tze.
\Cref{sec:method} explains how the spin function in SA-NNBF is chosen and constructed, and shows that the multi-determinant extension can represent any state in a chosen orbital space and spin sector.
\Cref{sec:Results} tests this extended ansatz on \ce{O2} and [2Fe-2S], including calculations of their low-lying excited states in chosen spin sectors.
\Cref{sec:Conclusion} summarizes the findings.

\section{Theory}\label{sec:Theory}

\subsection{Variational Monte Carlo (VMC) with neural quantum states (NQS)}\label{sec:VMC-NQS}
For a ground-state calculation,
the time-independent \Schrodinger equation in the occupation-number representation is
\begin{equation}
    \begin{aligned}
    	&\hat{H}\ket{\Psi} = E\ket{\Psi},\\
    	&\hat{H} = \sum_{pq}h_{pq}\hat{a}_{p}^\dagger\hat{a}_{q}+
    	\frac14\sum_{pqrs}\langle pq\|rs\rangle
    	\hat{a}_{p}^\dagger\hat{a}_{q}^\dagger
    	\hat{a}_{s}\hat{a}_{r}.
    \end{aligned}
\end{equation}
Here, $h_{pq}$ and $\langle pq\|rs\rangle$ are one-electron
and antisymmetrized two-electron integrals in the spin-orbital basis, respectively, and
$\hat{a}_p^{(\dagger)}$ is the fermionic
annihilation (creation) operator for the $p$-th
spin-orbital.
The NQS represents $|\Psi\rangle$ as
\begin{equation}
    |\Psi\rangle = \sum_{\bm n} \Psi(\bm n) |\bm n\rangle,
\end{equation}
where $\Psi(\bm n)=\langle \bm n\ket{\Psi}:\{0,1\}^{2K}\to\mathbb{F}$ is referred to as a (single) wavefunction ansatz.
Here, $\mathbb{F}=\mathbb{R}$ or $\mathbb{C}$, and $\bm n\in \{0,1\}^{2K}$ is an occupation-number vector (ONV) labeling a basis state in $\mathcal{S}$.
The number of spatial orbitals is $K$; the entries $0$ and $1$ denote an empty and an occupied spin orbital, respectively.
For a single ansatz, the energy of $\hat{H}$ can be estimated by averaging the local energy,
\begin{equation}
    \begin{aligned}
    	&E = \frac{\mel{\Psi}{\hat{H}}{\Psi}}{\langle \Psi|\Psi\rangle}
    	=
    	\langle E_{\rm loc}(\bm n)\rangle_{\bm n\sim P(\bm n)},\\
    	&E_{\rm loc}(\bm n)=\frac{\langle \bm n|\hat{H}|\Psi\rangle}{\langle \bm n|\Psi\rangle}
     =
    \sum_{\bm m}
    H_{\bm n\bm m}
    \frac{\Psi(\bm m)}{\Psi(\bm n)},
    \label{eq:eloc-sum}
    \end{aligned}
\end{equation}
where $H_{\bm n\bm m}\equiv\mel{\bm n}{\hat{H}}{\bm m}$, $P(\bm n) = |\Psi(\bm n)|^2/\sum_{\bm n}|\Psi(\bm n)|^2$ is the probability distribution,
and $\langle f(\bm n)\rangle_{\bm n\sim P(\bm n)}$ is defined as 
$\sum_{\bm n} P(\bm n)f(\bm n)$.

For a real variational parameter $\theta_i$, differentiating the energy expectation above gives\cite{2017-Becca}
\begin{align}
	g_i
    &=\frac{\partial E}{\partial\theta_i}\label{eq:vmc-parameter-update}\\
    &=2\operatorname{Re}
    \bigg\langle
    \frac{\partial\ln\Psi^*(\bm n)}{\partial\theta_i}
    \big(E_{\rm loc}(\bm n)-E\big)
    \bigg\rangle_{\bm n\sim P(\bm n)}.\nonumber
\end{align}
In VMC, the expectation is estimated from sampled ONVs. 
In simple stochastic gradient descent, 
the parameters are then updated as $\theta_i\leftarrow\theta_i-\eta g_i$, where $\eta>0$ is the learning rate.
AdamW\cite{2017-Kingma-Adam, 2019-Loshchilov-AdamW} modifies this gradient-based update, while stochastic reconfiguration (SR)\cite{1998-Sorella-Green} and its minimum-norm formulation (minSR)\cite{2024-Rende-Simple, 2024-Chen-Empowering} precondition the gradient using the overlap matrix of wavefunction derivatives.

For the electronic Hamiltonian, the local energy $E_{\rm loc}(\bm n)$ sums over single and double excitations of $\bm n$. In this work, we use the semistochastic local energy evaluation developed in our previous work.\cite{2025-Wu-Hybrid}
Specifically, a threshold $\epsilon>0$ decomposes the sum into a deterministic part and a stochastic part.
The deterministic part sums the excitations with $|H_{\bm n\bm m}|\ge\epsilon$.
The stochastic part samples $N_\epsilon$ configurations $\bm m'$ from $P_{\bm n}(\bm m')\propto|H_{\bm n\bm m'}|$, with $|H_{\bm n\bm m'}|<\epsilon$.
Then the estimator is
\begin{equation}
\begin{aligned}
    E_{\rm loc}(\bm n;\epsilon,N_\epsilon)
    &=
    \sum_{|H_{\bm n\bm m}|\ge\epsilon}
    H_{\bm n\bm m}
    \frac{\Psi(\bm m)}{\Psi(\bm n)}
    \\
    &\quad+
    \frac{1}{N_\epsilon}
    \sum_{\bm m'}
    \frac{H_{\bm n\bm m'}}{P_{\bm n}(\bm m')}
    \frac{\Psi(\bm m')}{\Psi(\bm n)}.
\end{aligned}
\label{eq:eloc-semistochastic}
\end{equation}
Its expectation equals \cref{eq:eloc-sum}, so the estimator is unbiased.
Appropriate choices of $\epsilon$ and $N_{\epsilon}$ can reduce the computational cost dramatically.

\subsection{Natural excited states (NES)}\label{sec:NES}

The NES method\cite{2024-Pfau-NES, 2025-Ma-NES, 2025-Hendry-GrassmannNES} transfers the problem of solving the lowest $M$ states of an $N$-electron system into solving the ground state of an expanded system containing $M$ non-interacting replicas of the original system,
\begin{equation}
	\hat{\mathcal{H}} 
	= \sum_{I=1}^{M} \hat{H}_I, 
\end{equation}
where $\hat{H}_I$ is the Hamiltonian of the $I$-th replica, and the wavefunction of the expanded system is defined as an determinant spanned by $M$ individual $N$-electron wavefunctions over the electron configuration of the $M$ replicas,
\begin{align}
    & \Psi_{\rm NES}(\bm{n}_{\rm NES}) = \operatorname{det}\bm{\Psi}_{\rm NES}(\bm{n}_{\rm NES}),
\end{align}
where
\begin{align*}
    & \bm{\Psi}_{\rm NES}(\bm{n}_{\rm NES}) \equiv
    \begin{bmatrix}
        \Psi_1(\bm n_1) & \cdots & \Psi_{M}(\bm n_1) \\
        \vdots & & \vdots \\
        \Psi_1(\bm n_{M}) & \cdots & \Psi_{M}(\bm n_{M}) \\
    \end{bmatrix}.
\end{align*}
Here $\bm{n}_{\rm NES}\equiv(\bm n_1, \bm n_2,...,\bm n_{M})$ with $\bm n_I$ being the occupation number vector of the $I$-th replica and each $\ket{\Psi_I}$ being a single $N$-electron state.\cite{2024-Pfau-NES, 2025-Ma-NES}
One could therefore minimize the NES energy,
\begin{align}
    E_{\rm NES} = \frac{\mel{\Psi_{\rm NES}}{\hat{\mathcal{H}}}{\Psi_{\rm NES}}}{\langle\Psi_{\rm NES}|\Psi_{\rm NES}\rangle},
\end{align}
where in the exact limit, the minimum would be the sum of the lowest $M$ eigenvalues of the original Hamiltonian, and the single-state functions $\{\Psi_I\}$ would span the corresponding low-energy subspace.
The individual variational energies and states are obtained by solving the generalized eigenvalue problem,
\begin{align}
    \bm{H}\bm{C}=\bm{S}\bm{C}\bm{\epsilon},\label{eq:nes-gevp}
\end{align}
where $H_{IJ}=\mel{\Psi_I}{\hat H}{\Psi_J}$, $S_{IJ}=\langle\Psi_I|\Psi_J\rangle$, and $\bm{\epsilon}=\operatorname{diag}(\epsilon_1^{\rm eig},\ldots,\epsilon_{M}^{\rm eig})$ with $\epsilon_{I}^{\rm eig}$ being the energy of the $I$-th state.
The $I$-th column of $\bm C$ defines the corresponding state,
\begin{align}
    \ket{\Psi_I^{\rm eig}}=\sum_{J=1}^{M}C_{JI}\ket{\Psi_J}.
\end{align}

When solving the NES problem using the VMC framework, the single-state functions $\{\Psi_I\}$ are parameterized and optimized, where the NES energy to be minimized is estimated through stochastic sampling by 
\begin{align}
    E_{\rm NES} = {\rm Tr}[\langle\bm{E}_{\rm loc}(\bm{n}_{\rm NES})\rangle_{\bm{n}_{\rm NES}\sim P(\bm{n}_{\rm NES})}],
\end{align}
where 
\begin{align}
    &\bm{E}_{\rm loc}(\bm{n}_{\rm NES})\nonumber\\
    =&
    \bm{\Psi}_{\rm NES}^{-1}(\bm{n}_{\rm NES})
    (\hat{\mathcal{H}}\bm{\Psi}_{\rm NES})(\bm{n}_{\rm NES})\label{eq:def-eloc}
\end{align}
is the matrix local energy with
\begin{align*}
    &(\hat{\mathcal{H}}\bm{\Psi}_{\rm NES})(\bm{n}_{\rm NES})\nonumber\\
    =&
    \begin{bmatrix}
        \langle\bm n_1|\hat{H}|\Psi_1\rangle & \cdots & \langle\bm n_1|\hat{H}|\Psi_M\rangle \\
        \vdots & & \vdots \\
        \langle\bm n_M|\hat{H}|\Psi_1\rangle & \cdots & \langle\bm n_M|\hat{H}|\Psi_M\rangle \\
    \end{bmatrix},
\end{align*}
and 
\begin{equation*}
    P(\bm{n}_{\rm NES}) = \frac{|\Psi_{\rm NES}(\bm{n}_{\rm NES})|^2}{\sum_{\bm{n}_{\rm NES}}|\Psi_{\rm NES}(\bm{n}_{\rm NES})|^2}.
\end{equation*}
In order to obtain the energy of each state after the variational optimization, one can estimate the matrix energy:
\begin{align}
    \bm{E}_{\rm mat} = 
	\langle\bm{E}_{\rm loc}(\bm{n}_{\rm NES})\rangle_{\bm{n}_{\rm NES}\sim P(\bm{n}_{\rm NES})},\label{eq:eloc(matrix)}
\end{align}
where it can be proved that $\bm{E}_{\rm mat} = \bm{S}^{-1}\bm{H}$, such as the eigenvalues of $\bm{E}_{\rm mat}$ are the desired $\{\epsilon^{\rm eig}_{I}\}$ according to \cref{eq:nes-gevp}; see the Supporting Information for more details.

\subsection{Wavefunction ans\"{a}tze}\label{sec:SA-NNBF}

\subsubsection{Vanilla neural network backflow (NNBF)}\label{sec:NNBF}

To smoothly transfer to the SA-NNBF framework via NNBF, we start with reviewing the well-established generalized Hartree\jux Fock (GHF) ansatz.
For an $N$-electron system in occupation-number representation, a set of occupied GHF orbitals can be encoded by a coefficient matrix $\mathbf{U}^{\rm GHF} \in \mathbb{F}^{2K\times N}$, where each column corresponds to an occupied GHF orbital represented as a combination of $2K$ basis spin orbitals, containing both spin-up and spin-down components.
Therefore, the transformed creation operators are
\begin{equation}\label{eq:Slater-operator}
	\hat{b}_j^\dagger = \sum_{p=1}^{2K} U^{\rm GHF}_{pj} \hat{a}_{p}^\dagger,\quad 
	j=1,\cdots, N,
\end{equation}
and the corresponding Slater determinant state is
\begin{equation}\label{eq:Slater-ansatz}
	\ket{\Psi^{\rm GHF}} =
	\prod_{j=1}^N \hat{b}_j^\dagger \ket{\vac},
\end{equation}
and one can derive that the corresponding amplitude in the original basis can be written as
\begin{equation}\label{eq:Slater-wavefunction}
	\langle \bm n \ket{\Psi^{\rm GHF}} = \det [\bm n\star \mathbf{U}^{\rm GHF}].
\end{equation}
Here, the symbol $\bm n\star$ represents the operation of selecting rows from the matrix on its right according to the occupied orbitals in $\bm n$.

The NNBF ansatz generalizes the GHF orbitals into backflow orbitals depending on the all-electron configuration,
and further expands the ansatz from a single determinant into a combination of multiple determinants.
It turns the constant matrix $\mathbf{U}^{\rm GHF}$ into a set of configuration-dependent matrices $\mathbf{U}^{\rm NNBF}_d(\bm n)$,
generated by a neural network with $\bm n$ as the input,
which gives
\begin{equation}\label{eq:NNBF}
	\langle \bm n \ket{\Psi^{\rm NNBF}}
	=
	\sum_{d=1}^{D}\det[\bm n\star \mathbf{U}^{\rm NNBF}_d(\bm n)],
\end{equation}
with $D$ being the number of determinants.

In this work, we use a multilayer perceptron (MLP) to implement $\mathbf{U}^{\rm NNBF}_d(\bm n)$,
which is specified by the number of layers $n_{\rm layer}$ and the number of hidden units $n_{\rm hidden}$.
Accordingly, we denote an NNBF model as NNBF($n_{\rm layer}$, $n_{\rm hidden}$, $D$).
Other architectures such as Transformer\cite{2025-Ma-Transformerbased, 2026-Gu-Solving} are also alternative, which are beyond the scope of this work.

\subsubsection{Spin-adapted neural network backflow (SA-NNBF)}\label{sec:SA-NNBF-for-fixed-spin-function}

Here, we briefly introduce the SA-NNBF ansatz \cite{2026-Li-SANNBF}.
We consider a system with $N$ electrons in $K$ spatial orbitals, with total spin $S$ and spin projection $M_{\rm s}$.
The electron spins can be coupled to form a spin function, denoted as $\Theta_{\sigma_1,\cdots,\sigma_N}$, which satisfies
\begin{equation}
    \hat{S}^2 \Theta = S(S+1)\Theta,\quad
    \hat{S}_z \Theta = M_{\rm s}\Theta, \label{eq:spin-eigen}
\end{equation}
and $\Theta$ can be expanded as a sum of products of one-electron spin functions,
\begin{align}
    \Theta_{\sigma_1,\cdots,\sigma_N} &= 
    \sum_{r=1}^R C_r\prod_{j=1}^N 
    \big(
        (s_r)_{1j} \alpha_{\sigma_j} +
        (s_r)_{2j} \beta_{\sigma_j}
    \big).\nonumber\\
    &=
    \sum_{r=1}^R C_r
    g_r(\sigma_1,\cdots,\sigma_N). \label{eq:sum-of-product}
\end{align}
Here, $\alpha_{\sigma_j}$ and $\beta_{\sigma_j}$ denote the spin-up and spin-down one-electron spin eigenfunctions for the spin variable $\sigma_j$, respectively\cite{2000-Helgaker-Molecular}.
The matrix $\mathbf{s}_r\in \mathbb{F}^{2\times N}$ gives the coefficients of these one-electron spin functions in the $r$-th product term, and $R$ denotes the number of product terms used to represent $\Theta$.
In this work, we give a detailed discussion on the alternative choices of $\Theta$ as well as the available implementations for \cref{eq:sum-of-product} in \cref{sec:method}.

For a fixed spin function $\Theta$, the SA-NNBF ansatz uses a neural network to produce spatial-orbital transformation matrices,
denoted as $\bar{\mathbf{U}}(\bar{\bm n})\in \mathbb{F}^{K\times N}$, 
with the spatial occupation number vector $\bar{\bm n}$ as the input. 
Here $\bar n_p=n_{p\alpha}+n_{p\beta}\in\{0,1,2\}$, where
$n_{p\alpha}$ and $n_{p\beta}$ denote the occupation numbers of the spin-up and spin-down orbitals associated with spatial orbital $p$, respectively.
With the spatial matrix and the spin function specified, the coefficient matrix $\mathbf{s}_r$ of the $r$-th product is combined with $\bar{\mathbf{U}}(\bar{\bm n})$ to form a spin-orbital coefficient matrix $\mathbf{U}_{r}(\bar{\bm n})\in\mathbb{F}^{2K\times N}$, whose components are
\begin{equation}\label{eq:urn-sannbf}
    [U_{r}]_{(p,\sigma),j}(\bar{\bm n})
    =
    \bar U_{pj}(\bar{\bm n}) \times (s_r)_{\sigma j}.
\end{equation}
This matrix depends only on the spatial occupation $\bar{\bm n}$, so configurations that differ only by a spin assignment share the same $\mathbf{U}_{r}(\bar{\bm n})$.
The amplitude on a configuration $\bm n$ is obtained by antisymmetrization: $\bm n\star$ selects the rows of the spin orbitals occupied in $\bm n$, and the determinant of those rows gives
\begin{equation}\label{eq:SA-NNBF-wavefunction}
    \Psi^{\text{\rm SA-NNBF}}(\bm n)
    =
    \sum_{r=1}^{R}C_r
    \det[\bm n\star \mathbf{U}_{r}(\bar{\bm n})].
\end{equation}
When \cref{eq:sum-of-product} represents $\Theta$ exactly, the state in \cref{eq:SA-NNBF-wavefunction} is an eigenfunction of both $\hat{S}^2$ and $\hat{S}_z$ for any spatial matrix $\bar{\mathbf{U}}(\bar{\bm n})$.
We denote this state by $\ket{\Psi^{\text{\rm SA-NNBF}}(\bar{\mathbf{U}}, \Theta)}$.
As a function of the trainable matrix $\bar{\mathbf{U}}$, the amplitude on configuration $\bm n$ is written $\Psi^{\text{\rm SA-NNBF}}[\bar{\mathbf{U}}](\bm n)$.

In analogy to NNBF and following Ref.~\citenum{2026-Li-SANNBF}, we simply choose an MLP to implement $\bar{\mathbf{U}}(\bar{\bm n})$ in SA-NNBF. 

\section{Spin functions in SA-NNBF}\label{sec:method}

As introduced above, the construction of an SA-NNBF contains two main parts: the spin part and the spatial part.
In this work, we give a brief discussion on the practical construction of the spin part, 
including the available choices of the spin eigenfunction $\Theta$ and the implementations of the sum-of-product decomposition in \cref{eq:sum-of-product}.
For the spatial part, we expand SA-NNBF with a combination of multiple sets of spatial orbitals to enhance its expressivity.
With such expansion, we prove that given a sufficiently large number of spatial orbital sets, an SA-NNBF with arbitrary spin function is able to reproduce any spin eigenfunctions.

\subsection{Choices of spin functions}\label{sec:choices-spin-func}

Since the electrons are identical particles, for any given target $S$ and $M_{\rm s}$, there exist many $\Theta$'s that satisfy \cref{eq:spin-eigen}\cite{1979-Pauncz-Spin},
where in SA-NNBF, the choice of $\Theta$ may affect both the accuracy 
and efficiency in practical VMC calculations.
Here, although a systematic numerical benchmark for all these choices is beyond the scope of this work, we briefly review some available methods to construct a spin eigenfunction.

The previous SA-NNBF work\cite{2026-Li-SANNBF} constructed the spin functions through genealogical coupling, a systematic scheme
where electrons are coupled one at a time and each electron raises or lowers the total spin by $1/2$.
For given $N$, $S$ and $M_{\rm s}$, the spin functions generated by this method form a complete orthogonal basis for the corresponding $N$-electron spin Hilbert subspace\cite{1979-Pauncz-Spin, 1974-Salmon-Genealogical, 2023-Ugandi-Recursive}. 
However, such scheme does not always offer an intuitive way to select one based on the chemical picture of a particular system,
such as systems containing multiple high-spin transition-metal centers, where the spins of each cluster are coupled as a whole rather than one electron at a time.

\begin{figure}[htbp]
    \centering
    \sbox{\OtwoPanelBox}{\includegraphics[width=\linewidth]{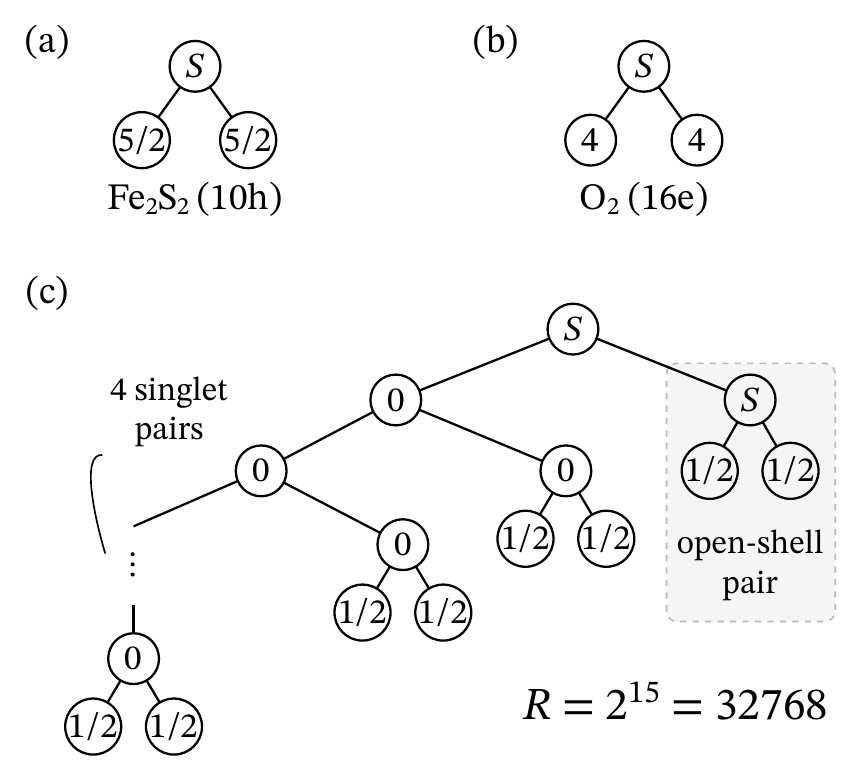}}%
    \makebox[0pt][l]{\hspace*{\dimexpr\wd\OtwoPanelBox*23/100\relax}%
        \raisebox{\dimexpr\ht\OtwoPanelBox*83/100\relax}[0pt][0pt]{\phantomsection\label{fig:results-tree-a}}}%
    \makebox[0pt][l]{\hspace*{\dimexpr\wd\OtwoPanelBox*75/100\relax}%
        \raisebox{\dimexpr\ht\OtwoPanelBox*83/100\relax}[0pt][0pt]{\phantomsection\label{fig:results-tree-b}}}%
    \makebox[0pt][l]{\hspace*{\dimexpr\wd\OtwoPanelBox/2\relax}%
        \raisebox{\dimexpr\ht\OtwoPanelBox*36/100\relax}[0pt][0pt]{\phantomsection\label{fig:results-tree-c}}}%
    \usebox{\OtwoPanelBox}
    \caption{Spin-coupling trees for (a) [2Fe-2S] in the ten-hole representation and (b,c) \ce{O2} in the 16-electron representation.
    In (c), four singlet pairs are omitted from the drawing.
    The $R$ in (c) is calculated from \cref{eq:R-SAAM}. 
    }
    \label{fig:results-tree}
\end{figure}

The SAAM approach of Ref.~\citenum{2025-Li-SAAM} provides another way 
to construct a spin wavefunction, which is through a coupling tree 
where the target total spin is obtained by hierarchically coupling the high-spin leaves along a binary tree, with each leaf at the highest possible spin according to its contained electrons. 
This scheme enables one to choose a coupling tree using chemical intuition,
such as for [2Fe-2S] where the chemical intuition suggests two local high-spin Fe(III) centers with each $S=5/2$.
In such system, we can use the hole-representation trick\cite{2016-Li-Hilbert, 2026-Li-SANNBF} to simplify the (20o,30e) active space model into a 10-hole (10h) model.
Following the two-center picture above, we divide the ten hole spins into two groups of five, couple each group to $S=5/2$, and then couple the two local spins to the target total spin $S$;
see \cref{fig:results-tree}\hyperref[fig:results-tree-a]{(a)}.

It is worth noting that different methods above can sometimes construct the same spin function.
A typical example of chemical interest is the valence bond (VB) spin function\cite{1979-Pauncz-Spin}:
\begin{align}
    \Theta^{\rm VB} & = \frac{1}{2^{N_{\rm pair}/2}}\Big[\prod_{k=1}^{N_{\rm pair}}(\alpha_{\sigma_{2k-1}}\beta_{\sigma_{2k}}-\beta_{\sigma_{2k-1}}\alpha_{\sigma_{2k}})\Big] \nonumber\\
    & \phantom{~\frac{1}{2^{N_{\rm pair}/2}}}
    \times \Big[\prod_{j=2N_{\rm pair}+1}^{N}\alpha_{\sigma_{j}}\Big],
    \label{eq:VB-spin-func}
\end{align}
where $N_{\rm pair} \equiv (N-2S)/2$ is the number of electron pairs.
In genealogical coupling scheme, \cref{eq:VB-spin-func} corresponds to the coupling path along which the first $2N_{\rm pair}$ electrons alternately raise and lower the intermediate total spin, while each remaining unpaired electron raises it\cite{1979-Pauncz-Spin}.
Equivalently, \cref{eq:VB-spin-func} can also be represented by a coupling tree as the example of \ce{O2} shown in \cref{fig:results-tree}\hyperref[fig:results-tree-c]{(c)}, which
couples each electron pair into an $S=0$ node and successively couples them together with the unpaired electrons to the target spin.

Beside the spatial\jux spin structure, another alternative perspective to construct a pure-spin many-body wavefunction is through spin projection (SP), where a spin-projection operator is applied to a symmetry-broken state\cite{1979-Pauncz-Spin, 1955-Lowdin-Quantuma, 2012-Jimenez-Hoyos-Projecteda, 2017-Li-SPMPS}.
Although the SP scheme does not explicitly introduce a spin eigenfunction, we can prove that such method is equivalent to the spatial\jux spin construction with an effective spin function corresponding to the simplest coupling tree; see the Supporting Information for more details.

\subsection{Sum-of-product decomposition of spin functions}\label{sec:Construction-of-spin-function}

In SA-NNBF, the spin function needs to be encoded into a sum-of-product format as shown in \cref{eq:sum-of-product} to apply antisymmetrization.
However, in most cases, the construction methods introduced in \cref{sec:choices-spin-func} do not directly give such format.
Therefore, a practical problem is how to decompose a spin function into such sum-of-product format.
Moreover, since the number of product terms $R$ could directly affect the computational cost, 
we desire to represent a spin function with as few terms as possible.

In principle, an optimal representation can be found through a projected CANDECOMP/PARAFAC (CP) decomposition\cite{2019-Kolda-Tensor, 2026-Li-SANNBF}.
Nevertheless, since the spin wavefunction is a high-order tensor,
a practical algorithm requires some special structures of the spin function to accelerate the evaluation of the loss function defined in the projected CP scheme. 
In Ref.~\citenum{2026-Li-SANNBF}, the spin functions are constructed by the genealogical coupling scheme, which could be written as a matrix product state (MPS), enabling efficient evaluation of the loss by tensor contraction. 

For the spin functions with a simple binary tree structure,
the original SAAM\cite{2025-Li-SAAM} provided an exact analytic decomposition, where the number of terms is at most
\begin{align}
    R_{\rm SAAM} = \prod_{l} (N_l + 1),
    \label{eq:R-SAAM}
\end{align}
in which the coefficient of a few terms might be coincidentally zero.
Here $N_l$ is the number of electrons in the $l$-th leaf of the binary tree, and the product excludes the leaf containing the most electrons.
Therefore, for a simple tree with only two leaves, such as \cref{fig:results-tree}\hyperref[fig:results-tree-a]{(a)} and \hyperref[fig:results-tree-b]{(b)}, $R_{\rm SAAM}$ is nearly half of the electron number, while for the tree in \cref{fig:results-tree}\hyperref[fig:results-tree-c]{(c)}, $R_{\rm SAAM}$ could scale exponentially.
Since such decomposition is based on Fourier transformation, its resulting coefficients are mostly complex numbers.
In this work, we improve this decomposition by
replacing the complex-valued Fourier transform\cite{2025-Li-SAAM} with real-valued interpolation, 
thereby making the resulting coefficients real; 
see the Supporting Information for details.
Although such improvement may still not reach the optimal decomposition from the perspective of minimizing $R$, such scheme is efficient enough for the proof-of-concept calculations in this work.

\subsection{An Example of a Spin Function}\label{sec:spin-function-example}

For molecular \ce{O2}, although the tree in \cref{fig:results-tree}\hyperref[fig:results-tree-b]{(b)} gives a much smaller $R$ than \cref{fig:results-tree}\hyperref[fig:results-tree-c]{(c)}, such coupling scheme does not seem to be consistent with the chemical intuition at a first glance, since the electrons in a molecule are usually believed to be mostly paired, rather than coupling into two high-spin clusters.
However, SA-NNBF with a tree like \cref{fig:results-tree}\hyperref[fig:results-tree-b]{(b)} can actually reproduce the paired-electron picture with a properly permuted spatial part. 

To simplify the algebra, we present the derivation for the triplet of a 6-electron system, while the results extend straightforwardly to other spin states as well as systems with more electrons, such as the 16-electron \ce{O2} molecule.
Moreover, we simplify the backflow spatial orbitals into fixed spatial orbitals, including two doubly occupied ones and two singly occupied ones:
\begin{align*}
	\bm{\Omega}_1 = (\varphi_a\varphi_a\varphi_b\varphi_b\varphi_c\varphi_d),
\end{align*}
where $\varphi_a$--$\varphi_d$ are spatial orbitals in the spatial product $\bm{\Omega}_1$.
One could verify that the SA-NNBF with $\bm{\Omega}_1$ as the spatial part and the VB spin wavefunction $\Theta^{\rm VB}$ in \cref{eq:VB-spin-func} as the spin part, is identical to a restricted open-shell Hartree\jux Fock (ROHF) determinant up to normalization:
\begin{align}
    \ket{\Psi^{\text{\rm SA-NNBF}}(\bm{\Omega}_1, \Theta^{\rm VB})} \propto \ket{\Psi^{\rm ROHF}}, \label{eq:ROHF}
\end{align}
where
\begin{align*}
    \ket{\Psi^{\rm ROHF}} \equiv \ket{\varphi_a\varphi_b\varphi_c\varphi_d\bar{\varphi}_a\bar{\varphi}_b},
\end{align*}
Such result is expected since the VB spin function and ROHF determinant correspond to a same paired-electron feature.

\begin{figure}[h]
	\centering
	\includegraphics{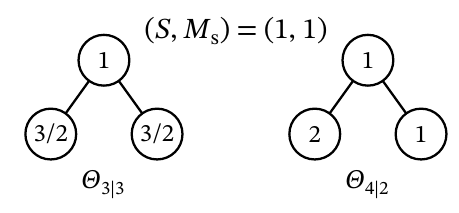}
	\caption{
    All the possible 2-leaf coupling tree with six electrons in the $(S,M_{\rm s})=(1,1)$ sector, giving the spin functions $\Theta_{3|3}$ and $\Theta_{4|2}$, respectively.
    }
	\label{fig:saam-tree}
\end{figure}

On the other hand, for the same system, all the possible 2-leaf simple trees are shown in \Cref{fig:saam-tree}, where the corresponding spin function are noted as $\Theta_{3|3}$ and $\Theta_{4|2}$.
The two trees give the following spin functions.
\begingroup
\allowdisplaybreaks[4]
\begin{align*}
    \Theta_{3|3}&={}
    \sqrt{\frac{1}{10}}(\alpha_1\alpha_2\alpha_3)\\*
    &\quad\times\big(
    \alpha_4\beta_5\beta_6+\beta_4\alpha_5\beta_6+\beta_4\beta_5\alpha_6
    \big)\\
    &-\frac{1}{3}\sqrt{\frac{2}{5}}
    \big(\alpha_1\alpha_2\beta_3+\alpha_1\beta_2\alpha_3+\beta_1\alpha_2\alpha_3\big)\\*
    &\quad\times\big(
    \alpha_4\alpha_5\beta_6+\alpha_4\beta_5\alpha_6+\beta_4\alpha_5\alpha_6
    \big)\\
    &+\sqrt{\frac{1}{10}}
    \big(\alpha_1\beta_2\beta_3+\beta_1\alpha_2\beta_3+\beta_1\beta_2\alpha_3\big)\\*
    &\quad\times(\alpha_4\alpha_5\alpha_6),\\
    \Theta_{4|2}&={}
    \frac{1}{\sqrt{60}}(\alpha_1\alpha_2)\\*
    &\quad\times\big(
    \alpha_3\alpha_4\beta_5\beta_6+\alpha_3\beta_4\alpha_5\beta_6+\beta_3\alpha_4\alpha_5\beta_6\\*
    &\qquad+\beta_3\alpha_4\beta_5\alpha_6+\beta_3\beta_4\alpha_5\alpha_6+\alpha_3\beta_4\beta_5\alpha_6\big)\\
    &-\frac{1}{4}\sqrt{\frac{3}{5}}
    (\alpha_1\beta_2+\beta_1\alpha_2)\\*
    &\quad\times\big(
    \alpha_3\alpha_4\alpha_5\beta_6+\alpha_3\alpha_4\beta_5\alpha_6\\*
    &\qquad+\alpha_3\beta_4\alpha_5\alpha_6+\beta_3\alpha_4\alpha_5\alpha_6\big)\\
    &+\sqrt{\frac{3}{5}}
    \beta_1\beta_2\alpha_3\alpha_4\alpha_5\alpha_6.
\end{align*}
\endgroup
Here, $\alpha_i$ and $\beta_i$ denote the one-electron spin functions on electron $i$.
Although \Cref{fig:saam-tree} seems to be different with the paired-electron picture, we can obtain exactly the same ROHF determinant by using a permuted spatial orbital set:
\begin{align*}
	\bm{\Omega}_2 = (\varphi_a\varphi_b\varphi_c\varphi_d\varphi_a\varphi_b),
\end{align*}
resulting in,
\begin{equation}
    \begin{aligned}
        \ket{\Psi^{\text{\rm SA-NNBF}}(\bm{\Omega}_2, \Theta_{3|3})} = & \frac{\sqrt{10}}{3} \ket{\Psi^{\rm ROHF}}, \\ 
        \ket{\Psi^{\text{\rm SA-NNBF}}(\bm{\Omega}_2, \Theta_{4|2})} = & \sqrt{\frac{5}{3}} \ket{\Psi^{\rm ROHF}}.
    \end{aligned}
    \label{eq:tree-to-rohf}
\end{equation}
In practice, since $\bar{\mathbf{U}}$ is obtained by optimization, \cref{eq:ROHF} and \cref{eq:tree-to-rohf} show that SA-NNBF is always able to reproduce the paired-electron picture since the ROHF state in within the representable space, regardless of whether $\Theta^{\rm VB}$, $\Theta_{3|3}$ or $\Theta_{4|2}$ is chosen as the spin function.

The results above in the 6-electron example could be straightforwardly extend to general systems, which makes it in general reasonable for us to use a 2-leaf simple coupling tree even in closed-shell systems.
Therefore, in this work, we use the tree in \cref{fig:results-tree}\hyperref[fig:results-tree-b]{(b)} rather than \hyperref[fig:results-tree-c]{(c)}, in order that the spin function can be decomposed into less terms while retaining the paired-electron chemical picture.

\section{SA-NNBF with multiple spatial-orbital sets}\label{sec:D-expansion}

Our previous work\cite{2026-Li-Representational} proved that given a sufficiently large hidden length $n_{\rm hidden}$, vanilla NNBF \cref{eq:NNBF} is universal with a single set of backflow orbitals, 
i.e. the NNBF is able to exactly represent an arbitrary target state.
However, SA-NNBF \cref{eq:SA-NNBF-wavefunction} does not have such property in general. 

To illustrate a potential limitation of the SA-NNBF ansatz with only a single set of backflow orbitals,
one could focus on the subspace of the configurations corresponding to a given spatial $\bar{\bm n}^0$,
where no matter how complicated is the neural network architecture, the SA-NNBF on these configurations share a fixed same $\bar{\mathbf{U}}(\bar{\bm n}^0) \equiv \bar{\mathbf{U}}^0$. Therefore, the SA-NNBF amplitudes within such subspace are completely determined by $\bm n$ and matrix $\bar{\mathbf{U}}^0$ according to \cref{eq:urn-sannbf} and \cref{eq:SA-NNBF-wavefunction}:
\begin{align*}
    \Psi^{\text{\rm SA-NNBF}}(\bm n) = \Psi^{\text{\rm SA-NNBF}}[\bar{\mathbf{U}}^0](\bm n), \,\, \forall \bm n \to \bar{\bm n}^0.
\end{align*}
To reproduce the exact wavefunction $\Psi^{\rm Exact}$, the SA-NNBF needs to satisfy the following set of equations:
\begin{align}
	\Psi^{\text{\rm SA-NNBF}}[\bar{\mathbf{U}}^0](\bm n) = \Psi^{\rm Exact}(\bm n), \,\, \forall \bm n \to \bar{\bm n}^0, \label{eq:sa-universal}
\end{align}
The equations in \cref{eq:sa-universal} are indexed by the configurations $\bm n$ whose spatial occupation is the chosen $\bar{\bm n}^0$.
Because $\ket{\Psi^{\rm Exact}}$ is a spin eigenstate, the right-hand sides $\Psi^{\rm Exact}(\bm n)$ for this set of $\bm n$ are linearly related.
The number of independent equations in the set equals the dimension of the spin subspace formed by the singly occupied orbitals in $\bar{\bm n}^0$, and this dimension still grows combinatorially with the number of singly occupied orbitals.\cite{1979-Pauncz-Spin}
Every equation in the set uses the entries of one matrix, $\bar{\mathbf{U}}^0$.
When $\bar{\bm n}^0$ places $N$ electrons in $N$ distinct spatial orbitals, only the $N$ rows of those orbitals enter the determinants, so $\bar{\mathbf{U}}^0$ contributes at most $N^2$ unknown entries.
For large $N$, the number of independent equations exceeds $N^2$.
\cref{eq:sa-universal} then has no solution for a general target state, and an SA-NNBF with a single spatial-orbital matrix cannot reproduce every state in the chosen $(N,S,M_{\rm s})$ sector.

To overcome the limitation above, in this work, we expand the SA-NNBF with multiple sets of spatial orbitals in analogy to
the $D>1$ form of \cref{eq:NNBF}:
\begin{equation}\label{eq:SA-NNBF-wavefunction-D}
	\Psi^{\text{\rm SA-NNBF}}_D(\bm n)
	=
	\sum_{d=1}^{D}
	\sum_{r=1}^{R}C_r
    \det[\bm n\star \mathbf{U}_{dr}(\bar{\bm n})],
\end{equation}
where
\begin{equation*}
    \begin{aligned}
    	[U_{dr}(\bar{\bm n})]_{(p,\sigma),j} = [\bar U_d(\bar{\bm n})]_{pj}(s_r)_{\sigma j},
    \end{aligned}
\end{equation*}
with $\{\bar{\mathbf{U}}_d(\bar{\bm n})\}$ produced by the neural network.
Following the notation for NNBF, we denote an SA-NNBF model configuration by SA-NNBF($n_{\rm layer},n_{\rm hidden},D$).

The expansion in \cref{sec:D-expansion} enables SA-NNBF to restore universality. In particular, we prove the following:
\begin{theorem}[Universality of SA-NNBF for fixed spin function]\label{thm:sannbf-exactness}
For any FCI state $\ket{\Psi_{(N,S,M_{\rm s})}^{\rm Exact}}$ in a fixed $(N,S,M_{\rm s})$ sector and any nonzero spin function $\Theta\in\mathcal{S}_{(N,S,M_{\rm s})}$, an SA-NNBF state with the fixed spin function $\Theta$ can exactly represent $\ket{\Psi_{(N,S,M_{\rm s})}^{\rm Exact}}$ when $D$ is sufficiently large.
\end{theorem}
A detailed proof is given in the Supporting Information.
\cref{thm:sannbf-exactness} also shows that the universality of SA-NNBF does not depend on the choice of $\Theta$, 
although in a practical calculation with a finite $D$, some specific choices may achieve higher accuracy or make
the corresponding fixed-spin-function ansatz easier to optimize.

\section{Results}\label{sec:Results}

In all calculations reported here, both NNBF and SA-NNBF use $n_{\rm layer}=1$.
For a given system, the number of trainable parameters $n_{\rm p}$ is then approximately proportional to $n_{\rm hidden}\times D$.
At the same $n_{\rm hidden}$ and $D$, NNBF has roughly twice as many trainable parameters as SA-NNBF since its backflow orbitals are spin orbitals rather than spatial orbitals.
Increasing $n_{\rm layer}$ may provide another route to higher accuracy; its effect remains to be investigated systematically in future.
Computational details are provided in the Supporting Information.

\subsection{Benchmarks of SA-NNBF}\label{sec:o2-nqs}

\begin{figure*}[t!]
	\centering
	\OtwoPanel{fig:o2-panel-a}{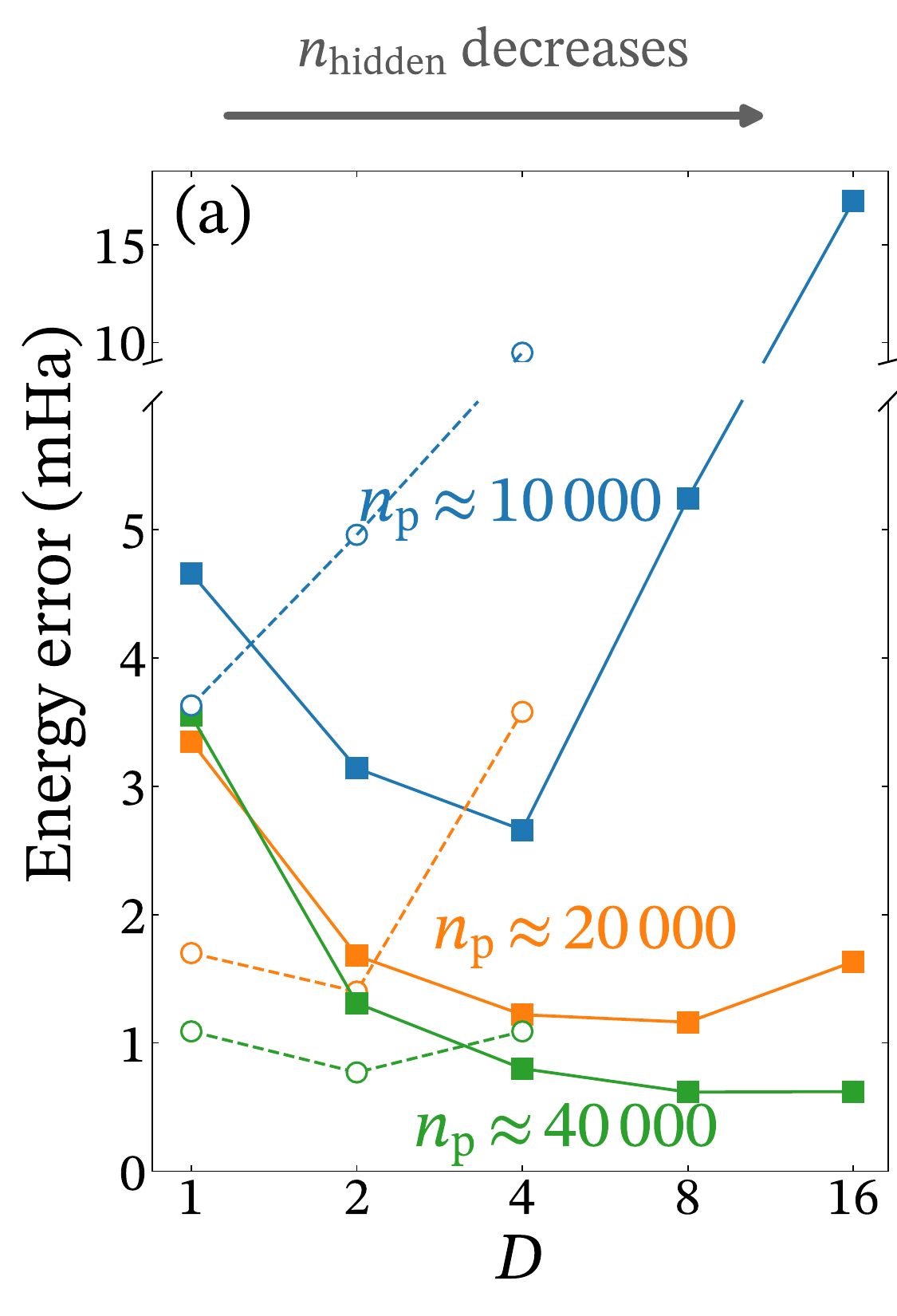}%
	\hfill
	\OtwoPanel{fig:o2-panel-b}{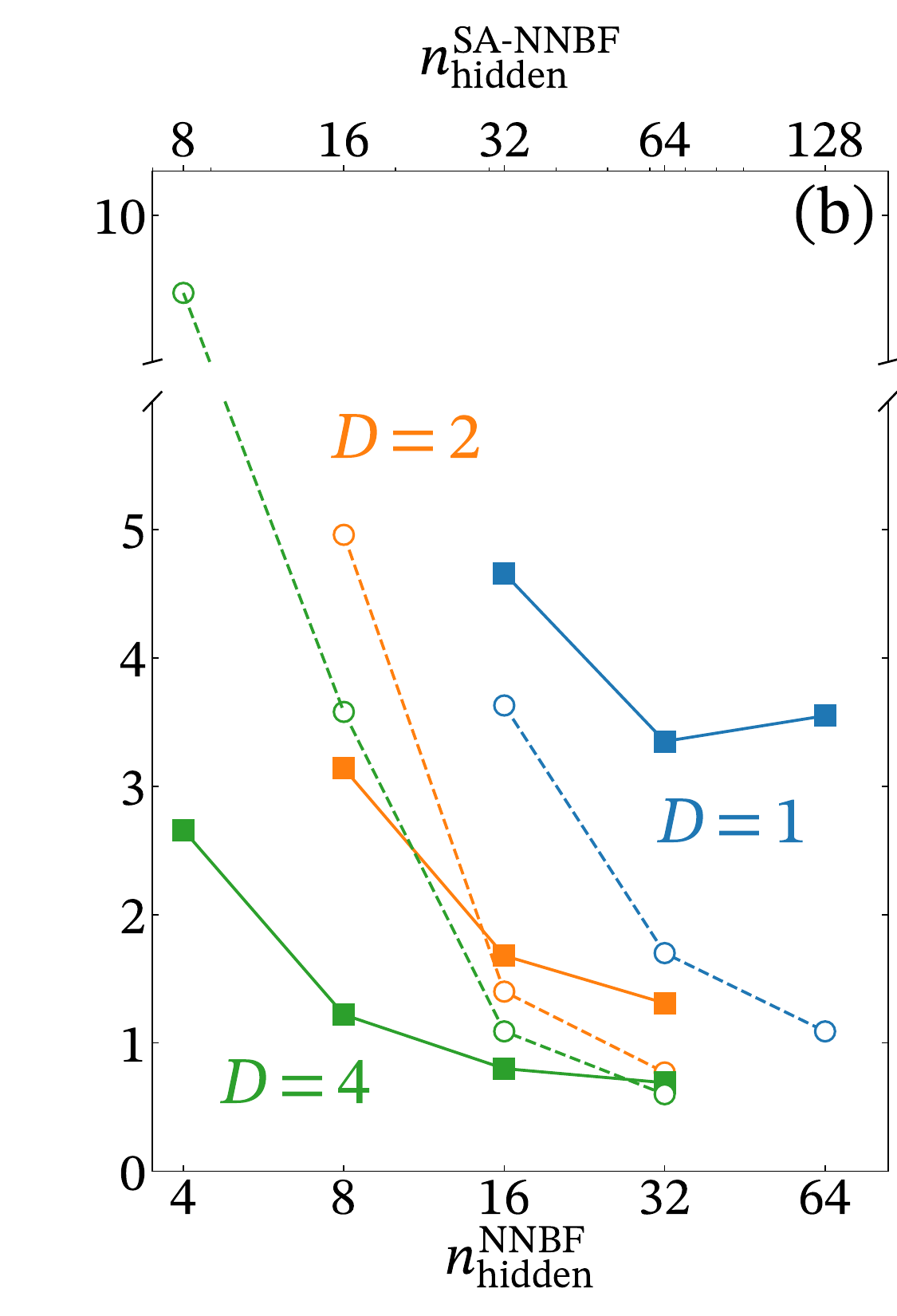}%
	\hfill
	\OtwoPanel{fig:o2-panel-c}{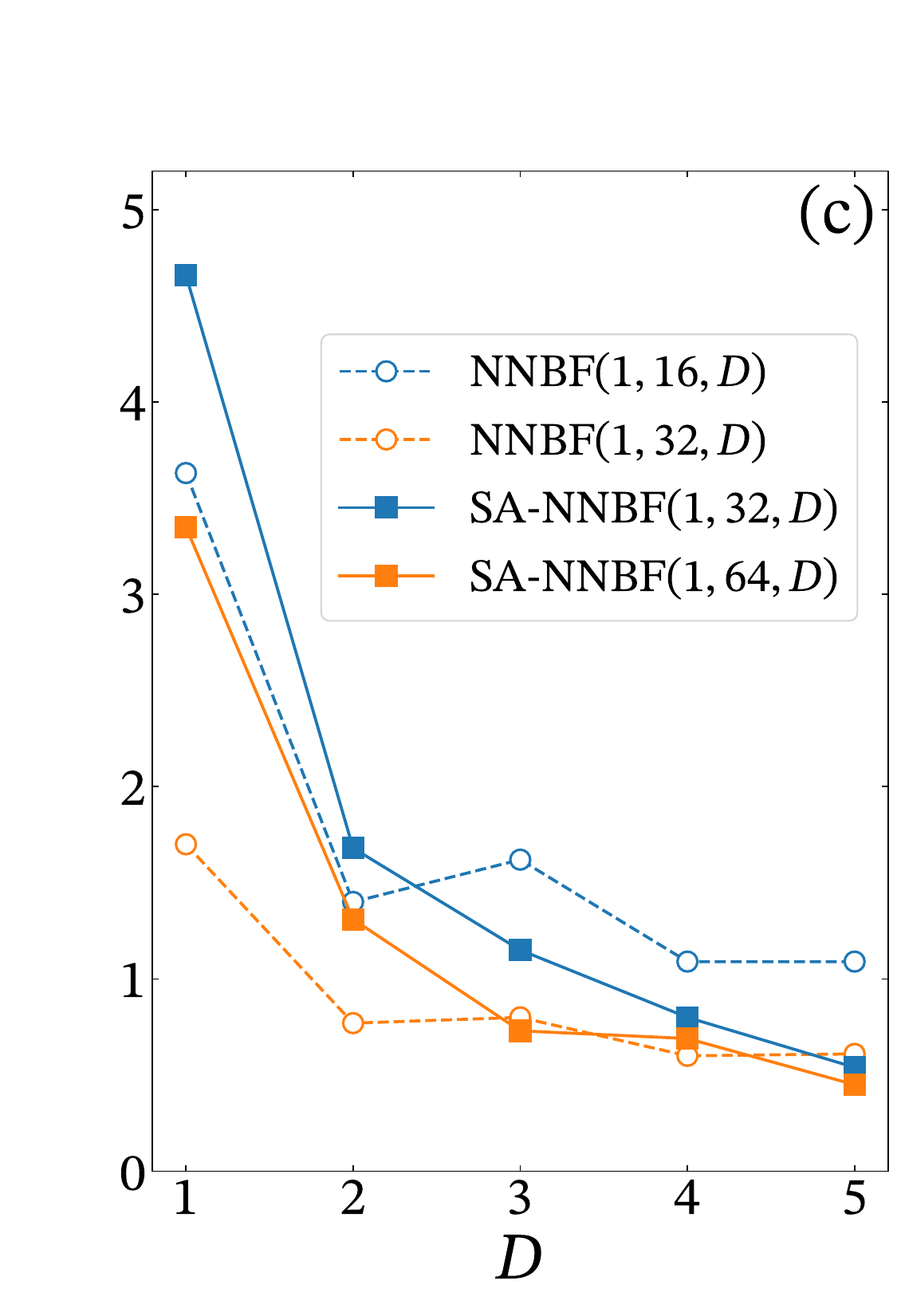}
	\caption{
    Energy errors of NNBF (dashed lines with open circles) and SA-NNBF (solid lines with filled squares) with different $n_{\rm hidden}$ or $D$ relative to FCI for the triplet ground state of \ce{O2} in the 6-31G basis with an (18o,16e) active space.
    For each vertically aligned pair of data points with the same color, there is $n_{\rm hidden}^{\text{\rm SA-NNBF}}=2\, n_{\rm hidden}^{\rm NNBF}$ in order that their parameter counts are comparable.
	(a) 
    Errors versus different parameter allocations. 
    Results of each color correspond approximately to the same parameter count, with $n_{\rm hidden}^{\rm NNBF}=16/D$, $32/D$ and $64/D$ at $n_{\rm p}\approx$ \num{10000}, \num{20000} and \num{40000}, respectively. For the same color, $n_{\rm hidden}^{\text{\rm SA-NNBF}}$ is twice $n_{\rm hidden}^{\rm NNBF}$. 
	(b) Errors versus hidden width at different fixed $D$'s
    where the lower and upper axes correspond to the hidden width for NNBF ($n_{\rm hidden}^{\rm NNBF}$) and SA-NNBF ($n_{\rm hidden}^{\text{\rm SA-NNBF}}$), respectively. 
	(c) Errors versus $D$ at different fixed NNBF hidden widths.
	}
	\label{fig:o2-nnbf-sannbf}
\end{figure*}

To investigate the performance of SA-NNBF, we test different allocations of $n_{\rm p}$ between $n_{\rm hidden}$ and $D$ and scaling strategies for SA-NNBF and compare them with NNBF.
We start with the triplet ground state of molecular \ce{O2}, as a simple case to benchmark different strategies.
\cref{fig:o2-nnbf-sannbf}\hyperlink{fig:o2-panel-a}{(a)} examines how the accuracy is affected as $D$ increases while $n_{\rm hidden}$ decreases to keep $n_{\rm p}$ approximately within a constant level.
For both NNBF and SA-NNBF, most of the curves have local minima rather than being entirely monotonic.
This suggests an optimal parameter allocation strategy at each $n_{\rm p}$ level, striking a balance between $n_{\rm hidden}$ and $D$.
However, the preferred strategies for the two ans\"atze differ substantially: the optimal $D$ values for SA-NNBF, typically $D\geq 4$, 
are consistently larger than those for NNBF, which are $1$ or $2$.

We 
further compare two different strategies of systematically scaling the model: 
increasing $n_{\rm hidden}$ at fixed $D$ in \cref{fig:o2-nnbf-sannbf}\hyperlink{fig:o2-panel-b}{(b)}, and increasing $D$ at fixed $n_{\rm hidden}$ in \cref{fig:o2-nnbf-sannbf}\hyperlink{fig:o2-panel-c}{(c)}.
In both panels, the energy error for each curve shows an overall decrease.
Within each pair of curves with the same color, the SA-NNBF curve (solid) shows a gentler slope and ends at a larger error than the NNBF curve (dashed) in panel \hyperlink{fig:o2-panel-b}{(b)}.
\hyperlink{fig:o2-panel-c}{(c)} shows the opposite pattern, where the SA-NNBF curve declines more steeply.  
Such a contrast shows a preference for increasing $D$ over $n_{\rm hidden}$ in SA-NNBF when scaling the model.

\begin{figure*}[t!]
	\centering
    \includegraphics[width=\textwidth,height=0.76\textheight,keepaspectratio]{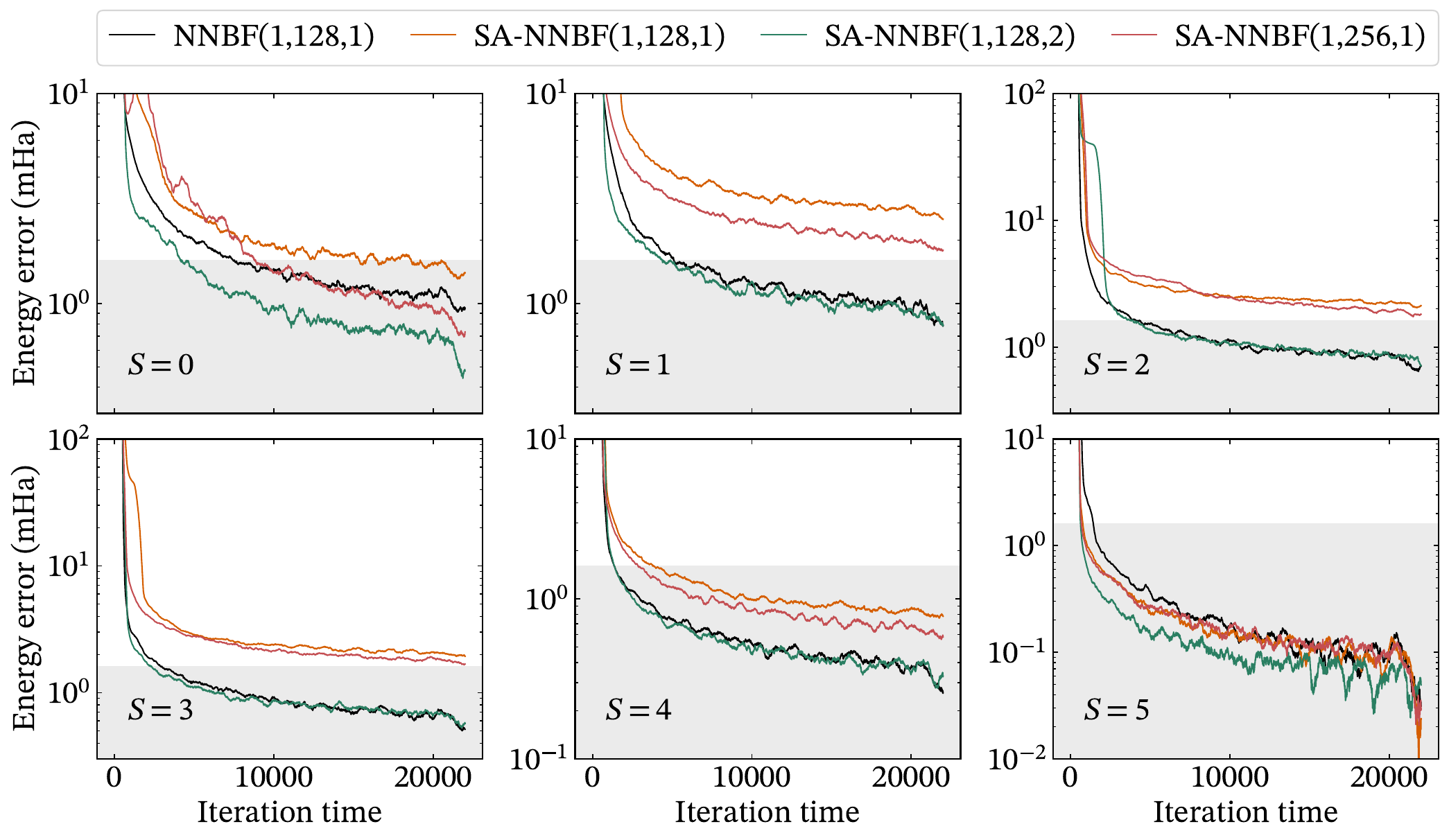}
	\caption{Optimization curves of the energy error relative to the corresponding FCI energy for the lowest [2Fe-2S] states with $S=0,\ldots,5$ and $M_{\rm s}=S$, obtained with vanilla NNBF(1,128,1) and SA-NNBF with different $D$ or $n_{\rm hidden}$. The active space is (20o,30e).
    The curves are smoothed using a $500$-step moving average.
    The shaded area represents the chemical accuracy ($<1$\,\unit{kcal/mol}).
    }
	\label{fig:fe2-nqs-sa-nnbf-bw}
\end{figure*}

To test SA-NNBF in more different spin sectors, we calculate the lowest states of [2Fe-2S] in the $S=0,1,...,5$ sectors respectively.
Since the energies of these states increase with $S$, 
we can directly use ground-state VMC to calculate the lowest state in each spin sector with both vanilla NNBF and SA-NNBF by simply setting $M_{\rm s}$ to the target $S$, without invoking NES.
As shown in \cref{fig:fe2-nqs-sa-nnbf-bw}, the SA-NNBF(1,128,1) failed to achieve chemical accuracy for half of the six states.
By increasing $D$ to $2$, resulting in SA-NNBF(1,128,2), SA-NNBF shows a substantial improvement, making its error consistently within \num{1}\,\unit{kcal/mol}.
This error is comparable with or even better than vanilla NNBF(1,128,1).
The two models have similar parameter counts.
By contrast, doubling $n_{\rm hidden}$ to obtain SA-NNBF(1,256,1) yields a final accuracy between those of SA-NNBF(1,128,1) and SA-NNBF(1,128,2), with only a modest improvement over SA-NNBF(1,128,1). 
Such results are consistent with the preference observed in \ce{O2}.

In summary, the \ce{O2} and [2Fe-2S] benchmarks show that, for the expanded SA-NNBF in this work, increasing $D$ is preferable to increasing $n_{\rm hidden}$ when systematically scaling the model to improve accuracy.
We believe this preference, compared with NNBF, is related to the universality results in \cref{sec:D-expansion}:
NNBF can achieve universality with either a sufficiently large $n_{\rm hidden}$ or a sufficiently large $D$, 
whereas SA-NNBF can do so only by increasing $D$ sufficiently, highlighting the more important role of $D$ in SA-NNBF.

\subsection{Spin-Adapted Excited-State Calculations}

The preceding benchmarks concern the lowest state in each spin sector.
We now use SA-NNBF to target excited states relative to the overall ground state, first with single-state VMC and then with NES.
For \ce{O2}, the lowest $S=0$ state is the first excited state of the molecule but the lowest state in the singlet sector.
SA-NNBF can therefore optimize it with a single-state VMC calculation, without an NES expansion.
As shown in \cref{tab:o2-sannbf-s0}, SA-NNBF(1,64,1) reaches an error of \num{1.46}\,\unit{mHa}, within the conventional chemical-accuracy threshold of \num{1.59}\,\unit{mHa}; increasing $D$ reduces the error to \num{0.46}\,\unit{mHa} at $D=5$.
By contrast, a vanilla NNBF calculation constrained only by $M_{\rm s}=0$ also includes the lower-energy $M_{\rm s}=0$ component of the triplet ground state and therefore cannot isolate this singlet by energy minimization alone.

\begin{table}[htpb]
\centering
\TableRuleSpacing
\setlength{\tabcolsep}{1pt}
\caption{Energies and absolute errors relative to FCI for the lowest singlet state of \ce{O2} from single-state SA-NNBF calculations. 
The spin function is chosen as the one represented by \cref{fig:results-tree}\hyperref[fig:results-tree-b]{(b)} with $S=0$.}
\label{tab:o2-sannbf-s0}
\begin{tabular*}{0.98\linewidth}{@{\extracolsep{\fill}}cccc@{}}
\hhline
model & $n_{\rm p}$ & \shortstack{energy\\(\unit{Ha})} & \shortstack{error\\(\unit{mHa})} \\
\hline
(1,64,1) & \num{19937} & \num{-149.74522(4)} & \num{1.46} \\
(1,64,2) & \num{38657} & \num{-149.74574(3)} & \num{0.94} \\
(1,64,3) & \num{57377} & \num{-149.74595(2)} & \num{0.72} \\
(1,64,4) & \num{76097} & \num{-149.74611(2)} & \num{0.56} \\
(1,64,5) & \num{94817} & \num{-149.74621(2)} & \num{0.46} \\
\hhline
\end{tabular*}
\end{table}

We further combine SA-NNBF with NES to calculate the lowest four triplets of \ce{O2}, including the ground state.
The four single ans\"{a}tze use the same fixed spin function shown in \cref{fig:results-tree}.
As shown in \cref{tab:o2-nes-results}, for each of the four triplets, SA-NNBF could achieve a high accuracy of \num{1}\,\unit{mHa} with $D=3$ for each single-state ansatz.
Meanwhile, with NNBF, one would need to include all the lowest 8-th states in an NES calculation to reach the fourth triplet, which would be far more expensive than the 4-state SA-NNBF calculation 
since constructing the NES wavefunction and local-energy matrices requires $M^2$ single-state evaluations per sampled configuration, in addition to the matrix-operation costs.

\begin{table}[htpb]
\centering
\TableRuleSpacing
\setlength{\tabcolsep}{10pt}
\caption{Energies and absolute errors relative to FCI for the four lowest triplet states of \ce{O2} from a 4-state SA-NNBF+NES calculation with four separately parameterized SA-NNBF(1,64,3) single-state ans\"{a}tze using the fixed spin function illustrated by \cref{fig:results-tree}\hyperref[fig:results-tree-b]{(b)} with $S=1$. The two $\mathrm{A}'\,{}^{3}\Delta_{\mathrm{u}}$ rows correspond to its different components obtained in NES calculations.}
\label{tab:o2-nes-results}
\begin{tabular*}{0.98\linewidth}{@{\extracolsep{\fill}}ccc@{}}
\hhline
state & energy (\unit{Ha}) & error (\unit{mHa}) \\
\hline
$\mathrm{X}\,{}^{3}\Sigma_{\mathrm{g}}^{-}$ & \num{-149.78801(5)} & \num{0.91} \\
$\mathrm{A}'\,{}^{3}\Delta_{\mathrm{u}}$ & \num{-149.55769(3)} & \num{0.55} \\
$\mathrm{A}'\,{}^{3}\Delta_{\mathrm{u}}$ & \num{-149.55728(5)} & \num{0.97} \\
$\mathrm{A}\,{}^{3}\Sigma_{\mathrm{u}}^{+}$ & \num{-149.55320(3)} & \num{0.64} \\
\hhline
\end{tabular*}
\end{table}

\begin{figure}[t]
	\centering
	\includegraphics[width=0.98\linewidth]{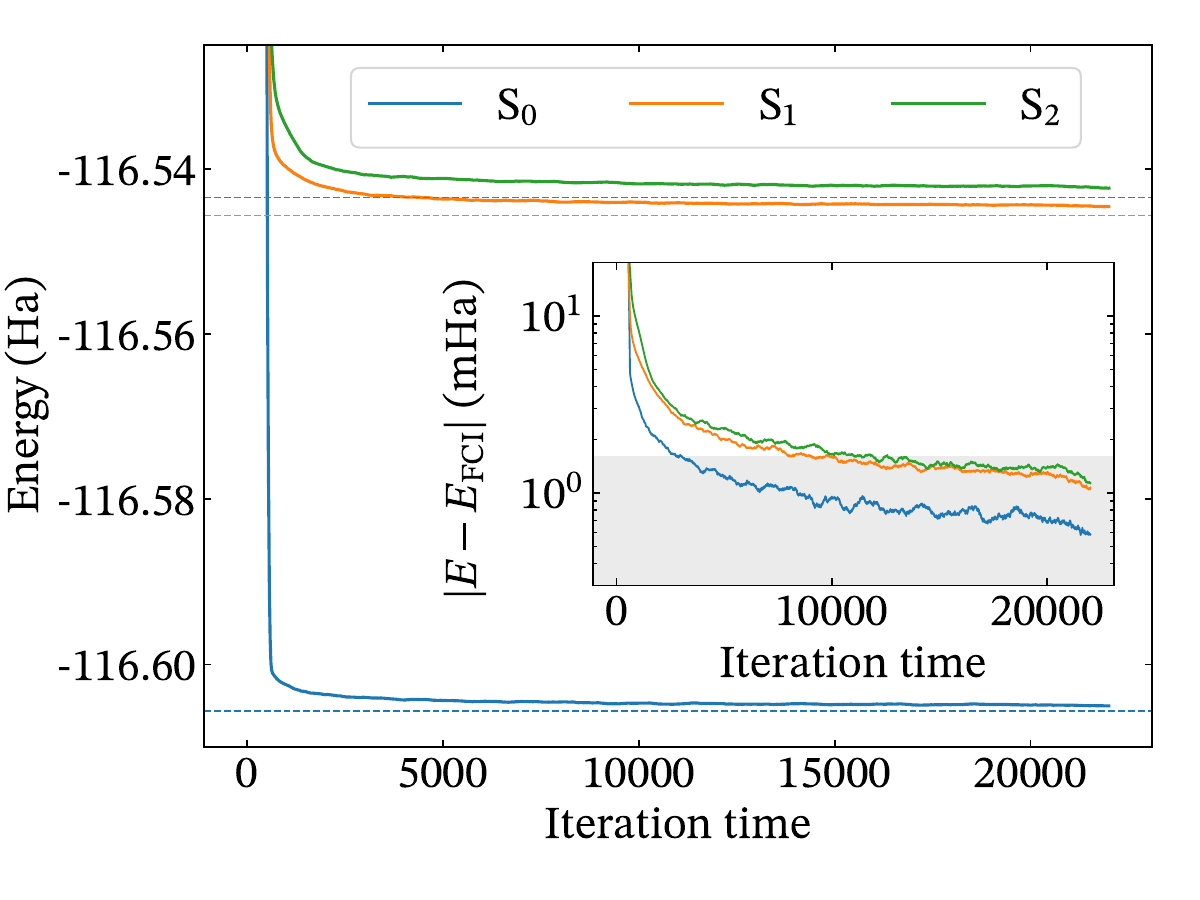}
	\caption{
        Optimization curves for the three lowest singlet states $( \mathrm{S}_0$, $\mathrm{S}_1, \mathrm{S}_2 )$ of [2Fe-2S] from a 3-state SA-NNBF+NES calculation. 
		The spin function is constructed using the same simple spin-coupling tree shown in \cref{fig:results-tree}\hyperref[fig:results-tree-a]{(a)} with $S=0$. 
		The optimization curves are smoothed using a $500$-step moving average. 
        The dashed lines indicate the FCI energies of the target states.
		The inset shows the absolute error of the same smoothed energies, for each state in \unit{mHa}; the shaded area represents chemical accuracy ($<1$\,\unit{kcal/mol}).
    }
	\label{fig:fe2-nes}
\end{figure}

Furthermore, we consider the [2Fe-2S] cluster, which is a typical example of correlated systems including many near-degenerate energy levels of different spins.
In seeking the three lowest singlet states $\mathrm{S}_0$, $\mathrm{S}_1$, and $\mathrm{S}_2$, 
there are over 30 higher-spin states lying between $\mathrm{S}_1$ and $\mathrm{S}_2$, 
all of which need to be included in a conventional NES calculation with NNBF to reach $\rm S_2$, 
making the computation unacceptably expensive.
However, as shown in \cref{fig:fe2-nes}, with SA-NNBF, we are able to compute the lowest three singlets with only a 3-state NES calculation.
Moreover, each of the three energy levels is optimized to chemical accuracy with a total parameter number of $\num{240267}$, 
which demonstrates the ability of SA-NNBF combined with NES to
provide accurate state-resolved energies for the ground and excited states of challenging systems.

\section{Conclusion}\label{sec:Conclusion}

In this work, we give a detailed discussion on SA-NNBF, including the choice and construction of spin functions. To overcome its potential limitations, we extended SA-NNBF to include $D$ sets of spatial orbitals and proved its universality for sufficiently large $D$.
Moreover, we performed a relatively systematic comparison of NNBF and SA-NNBF across the parameter allocation for the triplet ground state of \ce{O2} and the lowest states of [2Fe-2S] in different spin sectors.
Within the tested range, increasing $D$ improved the accuracy of SA-NNBF more effectively than increasing $n_{\rm hidden}$ alone, highlighting the central role of $D$ in SA-NNBF and providing guidance for future parameter choices.

Without any multi-state frameworks, we used SA-NNBF to calculate the lowest singlet of \ce{O2}, which is its first excited state.
Chemical accuracy was reached already at $D=1$, and increasing $D$ further improved the accuracy.
We further combined SA-NNBF with NES to calculate low-energy triplets of \ce{O2} and low-energy singlets of [2Fe-2S], achieving chemical accuracy in both cases.
Because the calculation is restricted to the target spin sector, it does not require the simultaneous calculation of many lower-energy intermediate states in other spin sectors.
While maintaining spin purity and energy accuracy, such spin restriction greatly reduces the computational resources required to obtain excited states of the target spin, making otherwise costly calculations feasible.

Although this work uses only MLP-based neural networks, NQSs with other architectures, such as the Transformer\cite{2023-Viteritti-Transformer, 2025-Shang-Solving, 2026-Gu-Solving, 2026-Rende-Transformer},
can also benefit from the spin-adaptation framework to accelerate the excited-state calculations by sufficiently reducing the involved states.
For more complex architectures, the optimal parameter allocations also deserve investigation, and the benchmark in this work could serve as a primary reference.
Besides, several avenues for improving the spin-adapted multi-state framework remain open, including selecting optimal spin functions for complex correlated systems, finding more compact spin-function decompositions, and adopting a multi-state formulation that scales more favorably with the number of states.
These could improve the efficiency and accuracy of this framework and enable us to calculate larger systems and therefore warrant further exploration in the future.

\section*{Acknowledgement}
This work was supported by the Quantum Science and Technology-National Science and Technology Major Project (2023ZD0300200) and the Fundamental Research Funds for the Central Universities.

\begingroup
\raggedright
\section*{Supporting Information Available}
\endgroup
The Supporting Information lists the symbols used here and proves the NES matrix local-energy identity.
It then develops a real-valued decomposition of binary-tree spin functions, constructs the spin function of the simplest coupling tree, and examines its connection to spin projection.
A proof of SA-NNBF exactness for a fixed spin function is followed by computational settings, FCI reference energies, and additional numerical results for [2Fe-2S] and \ce{O2}.

\bibliography{main}

\end{document}